\documentstyle[preprint,prd,aps,eqsecnum]{revtex}
\def\be{\begin{equation}}
\def\ee{\end{equation}}
\def\bea{\begin{eqnarray}}
\def\eea{\end{eqnarray}}
\def\beas{\begin{eqnarray*}}
\def\eeas{\end{eqnarray*}}
\def\no{\nonumber}
\def\a{\alpha}

\def\d{\delta}

\def\m{\mu}

\def\LLL{\Lambda}

\def\o{\omega}

\def\vph{\varphi}

\def\ps{\Psi}

\def\th{\theta}

\def\vA{{\bf A}}
\def\vB{{\bf B}}

\def\vE{{\bf E}}
\def\ve{{\bf e}}
\def\vf{{\bf f}}
\def\vj{{\bf j}}

\def\vl{{\bf l}}
\def\vp{{\bf p}}
\def\vP{{\bf P}}
\def\vr{{\bf r}}
\def\vR{{\bf R}}
\def\vu{{\bf u}}

\def\vx{{\bf x}}
\def\vy{{\bf y}}
\def\vz{{\bf z}}
\def\rt{({\vr},t)}

\def\vaa{\mbox{\boldmath $\alpha$}}
\def\vPP{\mbox{\boldmath $\Pi$}}
\def\na{\nabla}

\def\t{\widetilde}
\def\he{\hat{\ve}}

\def\hx{\hat{\vx}}
\def\hy{\hat{\vy}}
\def\hz{\hat{\vz}}
\def\bra{\langle}
\def\ket{\rangle}
\def\cd{\cdot}
\def\eq{\equiv}

\def\pa{\partial}
\def\ra{\rightarrow}

\def\cale{{\cal E}}
\def\calp{{\cal P}}

\def\ris{(\vr_1,\ldots\vr_N;t)}
\def\kin{\frac{\hbar^2}{2m_i}}
\def\pari{\frac{\pa}{\pa\vr_i}}
\def\qihc{\frac{q_i}{\hbar c}}

\def\ris{(\vr_1,\ldots,\vr_N;t)}
\def\kin{\frac{\hbar^2}{2m_i}}
\def\pari{\frac{\pa}{\pa\vr_i}}
\def\qihc{\frac{q_i}{\hbar c}}
\def\vcp{\calp\hspace{-10pt}\calp}
\def\vscp{\calp\hspace{-7pt}\calp}
\def\vsscp{\calp\hspace{-6pt}\calp}
\def\cales{{\raisebox{-3pt}{$\scriptstyle\cale$}}}

\begin{document}
\draft
\preprint{SNUTP 93-61}
\title{Charged Particle System in Uniform Magnetic and\\
Electric Fields: The Role of Galilean Transformation}
\author{Chanju Kim and Choonkyu Lee}
\address{Department of Physics and Center for Theoretical Physics\\
Seoul National University, Seoul, 151-742, Korea}
\maketitle
\begin{abstract}
Galilean transformation relates a physical system under mutually perpendicular
uniform magnetic and electric fields to that under uniform magnetic field
only. This allows a complete specification of quantum states in the former
case in terms of those for the latter. Based on this observation, we consider
the Hall effect and the behavior of a neutral composite system in the presence
of uniform electromagnetic fields.
\end{abstract}
\vspace{5mm}
\newpage
\section{INTRODUCTION}
Charged particle systems in the presence of uniform electromagnetic fields are
known to exhibit many remarkable features. Even in the simplest quantum
mechanical problem of a nonrelativistic charged particle moving in a uniform
magnetic field, we find the famous Landau energy levels [1] with infinite
degeneracy. Actually the infinite degeneracy, being a direct consequence of
the noncommutativity of the translation generators [2] in the presence of a
uniform magnetic field, is also to be found in many-particle systems with
translation-invariant interactions. When the external magnetic field is
sufficiently strong, the behavior of a physical system can be quite
extraordinary. The most spectacular related development in recent years is the
observation of the quantum Hall effect [3], which manifests itself as a series
of plateaus in the Hall resistance of materials containing two-dimensional
electron systems. A large number of literatures devoted to the theoretical
explanations of this effect has appeared since [4,5].

Analytical solutions to many body problems in the presence of background
electromagnetic fields are notoriously difficult. In this paper we describe
certain general consequences which stem solely from the Galilean transformation
property of underlying dynamical equations in the presence of a uniform
magnetic field. Specifically it will be shown that a complete specification of
any Galilean-invariant physical system in the presence of mutually
perpendicular uniform magnetic and electric fields can be made in terms of
those appropriate to the same system subject to a uniform magnetic field only.
The map connecting the two cases is the Galilean transformation, aside from
certain complications involving gauge transformation. Thanks to this map,
energy eigenvalues and energy eigenfunctions of the two cases are simply
related irrespectively of the details of interparticle interactions. The Hall
effect is in fact its most obvious realization in many electron systems, and
it has other application, e.g., on the behavior of a neutral composite system
such as a hydrogen atom or a neutron in the presence of uniform electromagnetic
fields. In the latter example, Galilean transformation accounts for the
intermixing between the center of mass motion and the Stark effect (in the
background of a uniform magnetic field). It also applies to the system of
non-relativistic anyons [6] with arbitrary Galilean invariant mutual
interaction.

In Ref. 2, Fubini also derived the formula for the change in the Landau energy
levels as an additional uniform electric field is introduced. But we find his
derivation physically less transparent since, in that paper, the Galilean
transformation behavior of the wave function is not addressed at all. Ours not
only clarifies the physical origin of the formula but also tells explicitly
how the wave functions change. The contents of this paper is quite elementary,
but, to our knowledge, not exploited much in the literature. We here hope to
convince readers that some of remarkable phenomena exhibited by matter in the
presence of uniform electromagnetic fields are just consequences of the
Galilean relativity principle.

This paper is organized as follows. Section II is a general exposition on the
realization of Galilean transformation in non-relativistic classical or
quantum systems, when a uniform background magnetic field is present. The
result is applied to simple one-body and two-body systems in the ensuing two
sections---classically in Section III and quantum mechanically in Section IV.
Especially the case of a net-neutral composite system, which has a rather
non-trivial dynamics, is dealt with in some detail. Section V contains a
summary and discussion. In the Appendix, the behavior of particle Green's
function under Galilean transformation is described.

\section{GALILEAN BOOST IN THE PRESENCE OF A UNIFORM MAGNETIC FIELD}
First, consider the case of classical mechanics. Given a uniform external
magnetic field $\vB=B\hz$ in some inertial frame, the equation of motion for a
system of $N$ Newtonian particles (with masses $m_i$ and charges $q_i$) will
read
\be  \label{2.1}
m_i\frac{d^2\vr_i}{dt^2}=\frac{q_i}{c}\frac{d\vr_i}{dt}\times\vB
        -\frac\pa{\pa\vr_i}V\quad (i=1,\ldots, N)\,.
\ee
Here, $V$ can be any Galilean-invariant interaction; but, for definiteness, we
set $V=\sum_{i<j}V(\vr_i-\vr_j)$, i.e., equal to the sum of two-body
interactions involving relative positions only. Then, in a primed coordinate
system obtained by the Galilean boost
\be
\vr'=\vr-\vu t\,,\quad\qquad t'=t\,,
\ee
we will have the equations of motion
\be  \label{2.3}
m_i\frac{d^2\vr'_i}{dt^2}=\frac{q_i}{c}\frac{d\vr'_i}{dt}\times\vB+q_i\vE
        -\frac\pa{\pa\vr'_i}\left(\sum_{i<j}V(\vr'_i-\vr'_j)\right)\,,
\ee
where $\vE\equiv\frac{\vu}c\times\vB$. Notice that we see also an external
electric field in the primed system. This shows that two problems---classical
dynamics in the presence of a uniform $\vB$-field and that in the presence of
uniform, mutually perpendicular, $\vB$- and $\vE$-fields---are simply related.
Explicitly, if $\vr_i=\vf_i(t)$ is any specific solution to (\ref{2.1}), then
\be   \label{2.4}
\vr'_i=\vf_i(t)-\vu t\,,\quad \qquad
       \left(\vu\equiv-c\frac{\vE\times\vB}{|\vB|^2}\right)
\ee
solves the system defined by (\ref{2.3}), i.e., the problem with an additional
uniform electric field $\vE$ (which is assumed to be perpendicular to $\vB$).
The piece $-\vu t$ in (\ref{2.4}) describes the famous Hall drift motion,
which is derived from the Galilean transformation behavior of the system
alone. Details of the interparticle interactions do not enter in this
discussion. Also, on the basis of (\ref{2.4}), one might think that the entire
effect of the additional electric field is to cause the constant drift of the
whole system only. That is not always so, for the separation of the center of
mass coordinates is not always trivial if a uniform magnetic field is
present [7]. See Sec. III on this.

We now turn to the corresponding quantum mechanical problem. The Hamiltonian
appropriate to the equations of motion (\ref{2.1}) is
\be
H=\sum_i\frac{1}{2m_i}\left[\vp_i-\frac{q_i}{c}\vA(\vr_i)\right]^2
     +\sum_{i<j}V(\vr_i-\vr_j)\,,
\ee
where the vector potential $\vA(\vr)$ should be chosen such that
$\na\times\vA(\vr)=B\hz$. Then the time development of a quantum state is
described by the wave function $\ps\ris$ satisfying the Schr\"{o}dinger
equation
\bea    \label{2.6}
i\hbar\frac\pa{\pa t}\ps\ris&=&(H\ps)\ris\no\\
 &=&\left\{-\sum_i\kin\left[\pari-i\qihc\vA(\vr_i)\right]^2
   +\sum_{i<j}V(\vr_i-\vr_j)\right\}\ps\ris\,.\no\\
   &&
\eea
On the other hand, in connection with the equations of motion (\ref{2.3}), we
can consider another Schr\"{o}dinger equation
\bea    \label{2.7}
i\hbar\frac\pa{\pa t}\ps'\ris
 &=&\left\{-\sum_i\kin\left[\pari-i\qihc\vA(\vr_i)\right]^2
   +\sum_{i<j}V(\vr_i-\vr_j)-\sum_iq_i\vE\cdot\vr_i\right\}\no\\
 & &\times\ps'\ris\,,
\eea
where $\vE$ can be any constant (electric field) vector subject to the
condition $\vE\cd\vB=0$. As in the classical motion, we assert that solutions
of these two problems are related by a suitable Galilean transformation modulo
gauge transformation. Explicitly, for any given solution $\ps\ris$ to
(\ref{2.6}), we have a corresponding solution to (\ref{2.7}) which has the
form
\be   \label{2.8}
\ps'\ris=e^{\frac{i}{\hbar}\a(\vr_1,\ldots,\vr_N,t;\vu)}
      \ps(\vr_1+\vu t,\ldots,\vr_N+\vu t;t)
\ee
with $\vu\eq-c\frac{\vE\times\vB}{B^2}$ and the (gauge-dependent) phase
$\a(\vr_1,\ldots,\vr_N,t;\vu)$ to be specified below.

In general, one can always express the vector potential leading to a constant
magnetic field $\vB$ by the form
\be  \label{2.9}
\vA(\vr)=\vB\times [\vr-(\vr\cd\hat\vu)\hat\vu]-\na\t\LLL(\vr)\,,
\ee
where $\hat\vu\eq\vu/|\vu|$ is assumed to be perpendicular to $\vB$ and
$\t\LLL(\vr)$ is an arbitrary gauge function. Note that, with
$\t\LLL(\vr)=0$, the given vector potential becomes invariant under the
Galilean transformation $\vr\ra\vr-\vu t$. If the vector potential has been
chosen such that $\t\LLL(\vr)$ may vanish, the phase
$\a(\vr_1,\ldots,\vr_N,t;\vu)$ in (\ref{2.8}) can in fact be identified with
the
usual 1-cocycle [8] appearing in the quantum mechanical realization of
Galilean transformations, namely,
\be  \label{2.10}
\a(\vr_1,\ldots,\vr_N,t;\vu)
 =-\sum_i\left\{m_i\vu\cd\vr_i+\frac{1}{2}m_i\vu^2t\right\}\quad\qquad
 (\mbox{for }\t\LLL(\vr)=0).
\ee
This may be checked by a direct computation; viz., if $\ps$ satisfies
(\ref{2.6}), the wave function $\ps'$ given by (\ref{2.8}) satisfies the
Schr\"{o}dinger equation of the form (\ref{2.7}) with
$\vE=\frac{\vu}{c}\times\vB$. [Note that, with $\vE\cd\vB=\vu\cd\vB=0$,
$\vE=\frac{\vu}c\times\vB$ is equivalent to
$\vu=-c\frac{\vE\times\vB}{|\vB|^2}$.] When one has the vector potential of the
general form (\ref{2.9}), the way to proceed is now obvious---consider the
Galilean
boost for suitably gauge-transformed wave functions $\ps$ and $\ps'$ (which
satisfy the appropriate Schr\"{o}dinger equations with the vector potential
given by the form invariant under the Galilean boost in question). This gives
rise to the following phase function:
\bea  \label{2.11}
\a(\vr_1,\ldots,\vr_N,t;\vu)
 =-\sum_i\left\{m_i\vu\cd\vr_i+\frac{1}{2}m_i\vu^2t\right\}&
  &-\sum_i\frac{q_i}{c}\left\{\t\LLL(\vr_i)-\t\LLL(\vr_i+\vu t)\right\}\no\\
 & & \hspace{10mm}\mbox{(general case).}
\eea

In the formula (\ref{2.8}) with the phase given by (\ref{2.11}), we have a
precise relation connecting general time-dependent wave functions of two
distinct systems, i.e., one in a uniform magnetic field $\vB$ and the other
subject to an additional constant electric field in the direction perpendicular
to $\vB$. This allows us to derive all pertinent informations concerning the
system under the external fields, say, $\vB=B\hz$ and $\vE=E\hy$ from those
for the system under $\vB=B\hz$ only, just by considering the Galilean
transformation with $\vu=-c\frac{E}{B}\hx$. For instance, a single-particle
Green's function in the presence of both $\vB$- and $\vE$-fields can be
written down using the Green's function defined in the presence of $\vB$ only.
This discussion, which further illuminates the meaning of the phase factor
$e^{\frac{i}{\hbar}\a}$, is relegated to the Appendix. Also, for the above
choice of field directions (as we shall assume below), the vector potential
invariant under the Galilean boost is assumed by the familiar Landau gauge,
i.e.,
\be
\vA(\vr)=-By\hx\quad\qquad \mbox{(Landau gauge).}
\ee

Of special importance will be the implications on the energy spectra and the
respective energy eigenfunctions. To study such, let
\be  \label{2.13}
\ps\ris=e^{-\frac{i}{\hbar}\cale t}
     \vph_\cales(\vr_1,\ldots,\vr_N)
\ee
represent a stationary solution to (\ref{2.6}). Then, in the Landau gauge, we
know on the basis of our formula (\ref{2.8}) that the wave function
\be  \label{2.14}
\ps'\ris=e^{-\frac{i}{\hbar}(\cale+\frac{1}{2}M\frac{c^2E^2}{B^2})t}
       e^{\frac{i}{\hbar}\frac{cE}{B}(m_1x_1+\cdots+m_Nx_N)}
       \vph_\cales\left(\vr_1-\frac{cEt}B\hx,\ldots,\vr_N-\frac{cEt}B\hx\right)
\ee
(here $M=m_1+m_2+\cdots+m_N$) should be a solution to the Schr\"{o}dinger
equation (\ref{2.7}). We now utilize the fact that, in the Landau gauge, we
have $[P_x,H]=0$ where $P_x\eq\sum_i\frac\hbar{i}\pari$ (i.e. equal to the
translation generator in the direction of the Galilean boost we make). Hence
we may require $\vph_\cales(\vr_1,\ldots,\vr_N)$ to be a simultaneous
eigenstate
of $H$ and $P_x$, with eigenvalues $\cale$ and $\calp_x$. This means that, in
(\ref{2.13}), we may substitute
\be  \label{2.15}
\vph_\cales(\vr_1,\ldots,\vr_N)=e^{\frac{i}{\hbar}\calp_xX}
  \bar\vph(Y,Z,\vl_1,\ldots,\vl_{N-1};\calp_x)
\ee
where $\vR\eq(X,Y,Z)=\frac{1}{M}\sum_im_i\vr_i$ represents the center of mass,
and $\vl_1,\ldots,\vl_{N-1}$ various relative coordinates whose values do not
depend on the choice of the spatial origin. We make the substitution also in
(\ref{2.14}), to obtain the wave function of the form
\be  \label{2.16}
\ps'=e^{-\frac{i}{\hbar}\cale't}\left[e^{\frac{i}{\hbar}\calp'_xX}
       \bar\vph(Y,Z,\vl_1,\ldots,\vl_{N-1};\calp_x)\right]\,,\hspace{20mm}\,
\ee
\vspace{-15mm}
\begin{mathletters}
\bea
\calp'_x&=&\calp_x+\frac{cE}{B}M\,,\label{2.17a}\\
\cale'&=&\cale+\frac{cE}{B}\calp_x+\frac{1}{2}M\frac{c^2E^2}{B^2}
    =\cale+\frac{cE}{B}\calp'_x-\frac{1}{2}M\frac{c^2E^2}{B^2}\,.\label{2.17b}
\eea
\end{mathletters}
These provide the connection between the energy eigenvalue spectra and energy
eigenfunctions of the two systems, and we emphasize that these relations apply
irrespectively of the details of interparticle interactions. Also, when the
center of mass dynamics cannot be trivially separated from the rest, the
simple change introduced in (\ref{2.16}) can alter the nature of the
stationary state significantly. We will illustrate this phenomenon with the
example of a two-body neutral atom in Sec. IV.

Note that the substitution (\ref{2.15}) can be made only in the Landau gauge.
If one wishes to work with the vector potential given by the general
expression (\ref{2.9}) (with $\hat\vu=\hx$), one may instead make the
substitution
\be  \label{2.18}
\vph_\cales(\vr_1,\ldots,\vr_N)=e^{-i\sum_i\qihc\t\LLL(\vr_i)}
      e^{\frac{i}{\hbar}\calp_xX}\bar\vph(Y,Z,\vl_1,\ldots,\vl_{N-1};\calp_x).
\ee
What we have in the right hand side here is the eigenstate of the conserved
translation generator in the $x$ direction, $\Pi_x$. In a general gauge,
$\Pi_x$ is specified as
\be  \label{2.19}
\Pi_x=\sum_i\left\{\frac\hbar{i}\frac{\pa}{\pa x_i}-\frac{q_i}cA_x(\vr_i)
     -\frac{q_i}c(\vr_i\times\vB)_x\right\}\,,
\ee
and the eigenvalue of this operator is $\calp_x$. Using this substitution with
our general formulas (\ref{2.8}) and (\ref{2.11}) then yields the expression
\be
\ps'=e^{-\frac{i}{\hbar}\cale't}e^{-i\sum_i\qihc\t\LLL(\vr_i)}
      e^{\frac{i}{\hbar}\calp'_xX}\bar\vph(Y,Z,\vl_1,\ldots,\vl_{N-1};\calp_x)
\ee
with $\cale'$ and $\calp'_x$ specified as before. Thus the only difference from
the Landau gauge case is the gauge transformation factor multiplying the
energy eigenfunctions, which is of course natural.

We may summarize our observation in the following way: a quantum system under
mutually perpendicular $\vB$- and $\vE$-fields is just the Galilean
transformation of the same system subject to the $\vB$-field only, and the
needed boost velocity $\vu=-c\frac{\vE\times\vB}{|\vB|^2}$ is precisely that of
the Hall drift motion. This connection can also be exhibited for the charge
and current densities. Assuming that the wave function $\ps$ has been properly
normalized, we may define the charge and current densities by
\begin{mathletters}
\bea
\rho\rt&=&\int\! d^3\vr_1\cdots d^3\vr_N\ps^*\ris
            \left(\sum_iq_i\d^3(\vr-\vr_i)\right)\ps\ris\,,\\
\vj\rt&=&\int\! d^3\vr_1\cdots d^3\vr_N\ps^*\ris
   \left(\sum_i\left(-\frac{i\hbar q_i}{2m_i}\right)
   \left[\pari-\loarrow{\pari}-\frac{2iq_i}{\hbar c}\vA(\vr)\right]\right)\no\\
   &&\hspace{27mm}\times\ps\ris\,,
\eea
\end{mathletters}
so that the current conservation, $\frac{\pa\rho}{\pa t}+\na\cd\vj=0$,
may hold.
Let $\rho(\vr)$ and $\vj(\vr)$ represent the appropriate quantities for the
stationary state given by (\ref{2.13}) (with the substitution (\ref{2.15})),
i.e., the quantities when the system is subject to the $\vB$-field only. Now,
if we evaluate the charge and current densities $(\rho'(\vr),\vj'(\vr))$ for
the wave function given by (\ref{2.16}), we immediately find the result
\be
\rho'(\vr)=\rho(\vr)\,,\qquad\quad\vj'(\vr)=\vj(\vr)+\rho(\vr)\frac{cE}B\hx\,.
\ee
Here the contribution $\rho(\vr)\frac{cE}B\hx$ is the Hall current density
which is seen in the presence of an additional electric field $\vE=E\hy$.
Note that this is an exact result as long as interparticle interactions
are Galilean invariant.

\section{APPLICATION TO SIMPLE CLASSICAL SYSTEMS}
We shall here illustrate the observation made in the previous section with
classical one-body and two-body systems. Field directions are taken as
$\vB=B\hz$ and $\vE=E\hy$. For a single charged particle, the equation of
motion
\be
m\frac{d^2\vr}{dt^2}=\frac{q}c\frac{d\vr}{dt}\times\vB+q\vE
\ee
can be solved easily. With $\vE=0$, it is a well-known cyclotron motion: viz.,
\be
\vr(t)=(v_zt+z_0)\hz+\vr_{0\perp}+K[\hx\cos(\o_ct-\th_0)+\hy\sin(\o_ct-\th_0)],
\ee
where $v_z$, $z_0$, $\vr_{0\perp}$, $K(>0)$, $\th_0$ are time-independent
parameters and $\o_c\eq\frac{qB}{mc}$ is the cyclotron frequency. Then,
with $\vE\neq0$, our recipe tells us that the general solution is simply
\be       \label{3.3}
\vr(t)=(v_zt+z_0)\hz+\vr_{0\perp}+K[\hx\cos(\o_ct-\th_0)+\hy\sin(\o_ct-\th_0)]
       + \frac{cE}Bt\hx\,.
\ee

For a two-body system, we may write the equations of motion as
\bea     \label{3.4}
m_1\frac{d^2\vr_1}{dt^2}&=&\frac{q_1}{c}\frac{d\vr_1}{dt}\times\vB+q_1\vE
                           -\frac\pa{\pa\vr_1}V\,,\no\\
m_2\frac{d^2\vr_2}{dt^2}&=&\frac{q_2}{c}\frac{d\vr_2}{dt}\times\vB+q_2\vE
                           -\frac\pa{\pa\vr_2}V\,,
\eea
where $V$ denotes some interparticle potential. Again the case of $\vE=0$ may
be considered first. We here restrict our attention to the cases for which the
center of mass coordinates $\vR=\frac{m_1\vr_1+m_2\vr_2}{m_1+m_2}$ can be
separated from the dynamics involving the relative position $\vr=\vr_1-\vr_2$.
There are two such situations [7]---viz., either
(i)$\frac{q_1}{m_1}=\frac{q_2}{m_2}=\frac{Q}M$ (with $Q=q_1+q_2$ and
$M=m_1+m_2$) or (ii) $q_1=-q_2=q$ (i.e. a net neutral system). We also take
the interparticle potential to be $V=\frac{k}2|\vr_1-\vr_2|^2$, so that we can
present closed-form solutions.

With $\frac{q_1}{m_1}=\frac{q_2}{m_2}=\frac{Q}{M}$, (\ref{3.4}) can be
rewritten as
\begin{mathletters}
\bea
&&M\frac{d^2\vR}{dt^2}=\frac{Q}c\frac{d\vR}{dt}\times\vB\,, \label{3.5a}\\
&&\frac{d^2\vr}{dt^2}=\frac{Q}{cM}\frac{d\vr}{dt}\times\vB-\frac{k}\m\vr\,,
                                    \label{3.5b}
\eea
\end{mathletters}
where $\m=\frac{m_1m_2}{m_1+m_2}$ is the reduced mass. The center-of-mass
dynamics is not different from the one particle motion discussed already. On
the other hand, using the variables $w=x+iy$ and $z$, (\ref{3.5b}) becomes
\bea
&&\frac{d^2z}{dt^2}+\frac{k}\m z=0\,,\no\\
&&\frac{d^2w}{dt^2}+i\frac{QB}{Mc}\frac{dw}{dt}+\frac{k}\m w=0\,.
\eea
These have the general solutions
\begin{mathletters}
\bea
z(t)&=&A_z\sin\left(\sqrt{\frac{k}{\m}}t+\th_0\right)\,,\label{3.7a}\\
w(t)&=&A_+e^{-i\a_+t}+A_-e^{-i\a_-t}\qquad\quad
  \left(\a_\pm=\frac{QB}{2Mc}
     \pm\sqrt{\frac{k}\m+\left(\frac{QB}{2Mc}\right)^2}\right), \label{3.7b}
\eea
\end{mathletters}
where $A_z$ and $\th_0$ are real while $A_\pm$ can be arbitrary complex
numbers. Using $\vr_\perp\eq(x,y)$ and setting
$\sqrt{\frac{k}\m+\left(\frac{QB}{2Mc}\right)^2}\eq\t\o_0$, (\ref{3.7b}) can
also be expressed by the form
\be  \label{3.8}
\vr_\perp(t)=a\he_1(t)\cos(\t\o_0t-\th_0)+b\he_2(t)\sin(\t\o_0t-\th_0)
\ee
with
\be
\he_1(t)=\cos\left(\frac{QB}{2Mc}t-\th_1\right)\hx
         -\sin\left(\frac{QB}{2Mc}t-\th_1\right)\hy,\qquad
\he_2(t)=\hz\times\he_1(t),
\ee
where $a$, $b$, $\th_0$ and $\th_1$ are some real constants. Note that, for
$B=0$, $\he_1$ and $\he_2$ become space-fixed and the orbit becomes elliptical.
But, for $B\neq0$, the axes of the ellipse also undergo uniform rotation with
frequency $\frac{QB}{2Mc}$. [This is in accord with the classical Larmor
theorem.] Having the additional electric field $\vE=E\hy$ does not affect the
relative dynamics; its entire effect is to add the drift motion term
$\frac{cE}Bt\hx$ in $\vR(t)$ (just as in (\ref{3.3})).

More interesting is the case of $q_1=-q_2=q$. Here, employing the center of
mass and relative coordinates, we can cast (\ref{3.4}) for $\vE=0$ as
\begin{mathletters}
\bea
M\frac{d^2\vR}{dt^2}&=&\frac{q}c\frac{d\vr}{dt}\times\vB\,, \label{3.10a}\\
\m\frac{d^2\vr}{dt^2}&=&
 \frac{q}c\frac{m_2-m_1}{m_1+m_2}\frac{d\vr}{dt}\times\vB
 +\frac{q}c\frac{d\vR}{dt}\times\vB-k\vr\,.\label{3.10b}
\eea
\end{mathletters}
{}From (\ref{3.10a}), we notice the existence of the first integrals
\begin{mathletters}
\bea
&&M\frac{dZ}{dt}=\calp_z\hspace{4mm}(\mbox{:const.})\,,\label{3.11a}\\
&&M\frac{d\vR_\perp}{dt}-\frac{q}c\vr_\perp\times\vB=\vcp_\perp
           \hspace{4mm}(\mbox{:const.})\,,\label{3.11b}
\eea
\end{mathletters}
and, using these relations, (\ref{3.10b}) can be made as equations for $\vr$
only:
\begin{mathletters}
\bea
\frac{d^2z}{dt^2}&=&-\frac{k}\m z\,,\label{3.12a}\\
\frac{d^2\vr_\perp}{dt^2}&=&
    \frac{q}c\frac{m_2-m_1}{m_1+m_2}\frac{d\vr_\perp}{dt}\times\vB
   -\left(\frac{k}\m+\frac{q^2B^2}{c^2m_1m_2}\right)\vr_\perp
   +\frac{q}{c m_1m_2}\vcp_\perp\times\vB. \label{3.12b}
\eea
\end{mathletters}
[In this paper, the subscript $\perp$ is attached to indicate any vector which
is defined to be perpendicular to the direction of $\vB$]. Due to
(\ref{3.11b}), the center of mass motion is not completely independent from the
relative motion in the present case. Motion in the direction of $\vB$ requires
no explanation. On the other hand, to analyze the transverse motion, we find it
convenient to introduce the displaced relative position
\be  \label{3.13}
\vr'_\perp=\vr_\perp-\frac{q}{cM}\vaa_\perp\times\vB,\qquad\quad
  \left(\vaa\eq\frac{\vcp_\perp}{k+(q^2B^2/c^2M)}\right).
\ee
then, (\ref{3.11b}) and ({\ref{3.12b}) read
\begin{mathletters}
\bea
M\frac{d\vR_\perp}{dt}
 &=&\frac{q}c\vr'_\perp\times\vB+k\vaa_\perp\,, \label{3.14a}\\
\m\frac{d^2\vr'_\perp}{dt^2}&=&
 \frac{q}c\frac{m_2-m_1}{m_1m_2}\frac{d\vr'_\perp}{dt}\times\vB
 -\frac{1}\m\left(k+\frac{q^2B^2}{c^2M}\right)\vr'_\perp\,.\label{3.14b}
\eea
\end{mathletters}
Equation (\ref{3.14b}) has the same form as (\ref{3.5b}), and hence its general
solution is provided by the expression (\ref{3.8}) except for the fact that we
here have
$\t\o_0=\sqrt{\frac1\m(k+\frac{q^2B^2}{c^2M})
 +\frac{q^2B^2(m_1-m_2)^2}{4c^2m_1^2m_2^2}}$ and the axes $\he_1(t)$,
$\he_2(t)$ rotate with frequency $\frac{qB(m_2-m_1)}{2m_1m_2}$. Using this
solution for $\vr'_\perp(t)$ in (\ref{3.14a}) one can determine the center of
mass motion as well and the result will clearly have the form
\be    \label{3.15}
\vR_\perp(t)=\vR_0+\frac{k}{M}\vaa_\perp t
  +\left(\matrix{\mbox{\raisebox{-1.5mm}{piece oscillating about}}\cr
   \mbox{\raisebox{1.5mm}{the zero vector}}}\right).
\ee
Here, $\vaa_\perp$ can be any constant vector.

There is a certain noteworthy point with the above solution. If
$\vaa_\perp=0$, both $\vr_\perp(t)$ and $\vR_\perp(t)$ will oscillate about
zero. But, when $\vaa_\perp\neq0$ or equivalently the center of mass has
non-zero average velocity $\bra\dot\vR_\perp\ket=\frac{k}M\vaa_\perp$, the
(transverse) relative vector $\vr_\perp(t)$ will oscillate about the average
value
\be   \label{3.16}
\bra\vr_\perp\ket=\frac{q}{ck}\bra\dot\vR_\perp\ket\times\vB\,.
\ee
This shows that the composite acquires an average electric dipole moment
$\frac{q^2}{ck}|\bra\dot\vR_\perp\ket\times\vB|$ in the direction
perpendicular to both $\vB$ and $\bra\dot\vR_\perp\ket$. This behavior is
easily understood for the special (non-oscillating) solution given by
$\vr_\perp(t)=\frac{q}{cM}\vaa_\perp\times\vB$ and
$\vR_\perp(t)=\vR_0+\frac{k}{M}\vaa_\perp t\;$---then, (\ref{3.16}) is nothing
but the condition that the Lorentz forces acting on individual charges should
balance the attractive interparticle forces. Alternatively, one may
contemplate on making the Galilean boost to the frame in which the very
center of mass is at rest. The above phenomenon can then be seen as the
Stark-type effect due to the electric field thus generated. The latter view is
also closely related to the discussion that follows.

When this neutral system is subject to an additional electric field
$\vE=E\hy$, the corresponding general solution can be now written down with no
effort. On the basis of (\ref{2.4}), all that the $\vE$-field does is to
introduce an extra uniform motion term $c\frac{\vE\times\vB}{B^2}t$ in
$\vR(t)$; namely in the expression (\ref{3.15}), the piece
$\frac{k}M\vaa_\perp t$ gets effectively replaced by
$(\frac{k}M\vaa_\perp+c\frac{\vE\times\vB}{B^2})t$, while the expression for
$\vr_\perp(t)$ is unaffected by the presence of the $\vE$-field. We now have
the formula
$\bra\dot\vR_\perp\ket=\frac{k}M\vaa_\perp+c\frac{\vE\times\vB}{B^2}$ for the
average center of mass velocity, which implies that the relationship between
$\bra\vr_\perp\ket$ and $\bra\dot\vR_\perp\ket$ in the presence of the
$\vE$-field also reads
\be   \label{3.17}
\bra\vr_\perp\ket=\frac{q}{ck}\bra\dot\vR_\perp\ket\times\vB
                   +\frac{q}k\vE\,.
\ee
Evidently, this has an obvious explanation in terms of the force balance
again. The average electric dipole moment $\bra\vp_e\ket=q\bra\vr_\perp\ket$
points in the direction of the $\vE$-field (the usual Stark effect) only when
$\bra\dot\vR_\perp\ket=0$; but with a non-zero center of mass velocity, it will
assume the direction of $q\vE'$ with
$\vE'=\vE+\frac1c\bra\dot\vR_\perp\ket\times\vB$. In particular, if the average
center of mass velocity is equal to $c\frac{\vE\times\vB}{B^2}$ (this amounts
to $\vaa_\perp=0$), we find $\bra\vr_\perp\ket=0$, i.e., the system has zero
electric dipole moment; this is the result of cancellation between the force
due to the $\vE$-field and the Lorentz force. Although we have only considered
a specific, explicitly solvable, example here, it is clear that analogous
phenomena should be exhibited by any neutral composite system.

\section{APPLICATION TO SIMPLE QUANTUM SYSTEMS}
In this section we will give quantum mechanical discussions for the same
one-body and two-body systems that we considered classically in Section III.
Here our main concern will be directed to the effects on the energy
eigenstates, as an additional electric field is introduced into the system
already subject to a uniform magnetic field. Conventions for the field
directions are as in Section III.

The Hamiltonian for a single charged particle is
\be   \label{4.1}
H=\frac1{2m}\left[\vp-\frac{q}c\vA(\vr)\right]^2-qEy\,,\qquad
      (\na\times\vA=B\hz)\,.
\ee
For $E=0$ (i.e., zero electric field), the energy eigenstates are familiar
Landau levels [1]. Especially in the Landau gauge with $\vA=-By\hx$, $p_x$ and
$p_z$ commute with $H$. So it suffices to diagonalize the Hamiltonian with
$p_x(p_z)$ replaced by its eigenvalue $\calp_x(\calp_z)$, and one finds
essentially a one-dimensional harmonic oscillator problem with
\be   \label{4.2}
H=\frac1{2m}p_y^2+\frac12m\o_c^2\left(y+\frac{c\calp_x}{qB}\right)^2
  +\frac1{2m}\calp_z^2\,,\qquad\quad
  \left(\o_c\eq\left|\frac{qB}{mc}\right|\right).
\ee
{}From this, it follows that the energy eigenfunction corresponding to the
eigenvalue
$\cale_{n,\calp_x,\calp_z}=\hbar\o_c(n+\frac12)+\frac{\calp_z^2}{2m}$
($n=0,1,2,\ldots$) is
\be   \label{4.3}
\vph_{n,\calp_x,\calp_z}^L(\vr)
 =(\mbox{const.})e^{\frac{i}\hbar(\calp_xX+\calp_zZ)}
 e^{-\frac{m\o_c}{2\hbar}(y+\frac{c\calp_x}{qB})^2}
 H_n\left(\sqrt{\frac{m\o_c}\hbar}\left(y+\frac{c\calp_x}{qB}\right)\right)\,,
\ee
where $H_n$ is the $n$-th degree Hermite polynomial. Infinite degeneracy of
the Landau level is manifest in the fact that the energy eigenvalue has no
dependence on the quantum number $\calp_x$ at all. When there is a non-zero
electric field $\vE=E\hy$, we then make use of (\ref{2.16}), ({\ref{2.17a})
and ({\ref{2.17b}) to obtain the corresponding exact energy eigenstates:
\bea      \label{4.4}
\cale_{n,\calp_x,\calp_z}&=&\hbar\o_c\left(n+\frac12\right)+\frac{cE}B\calp_x
            -\frac{mc^2E^2}{2B^2}+\frac{\calp_z^2}{2m}\,,\no\\
\vph_{n,\calp_x,\calp_z}^L
 &=&\mbox{(const.)}e^{\frac{i}\hbar(\calp_xX+\calp_zZ)}
    e^{-\frac{m\o_c}{2\hbar}(y+\frac{c\calp_x}{qB}-\frac{mc^2E}{qB^2})^2}
    H_n\left(\sqrt{\frac{m\o_c}\hbar}\left(y+\frac{c\calp_x}{qB}
          -\frac{mc^2E}{qB^2}\right)\right)\,.\no\\
 &&
\eea
Infinite degeneracy of the Landau level is lifted by the electric field.

As we explained in Section II, one may work in other gauges as well although,
given the electric field $\vE=E\hy$, the above Landau gauge treatment is
somewhat simpler. Take, for instance, the symmetric gauge with
$\vA=-\frac12\vr\times\vB$. For $\vE=0$, one normally considers the
simultaneous
eigenstates of $H$, $p_z$ and $L_z=\frac\hbar{i}(xp_y-yp_x)$ in this gauge.
But, when one has in mind subjecting the system to the electric field
$\vE=E\hy$, the more appropriate are the eigenstates of $H$, $p_z$ and
$\Pi_x=p_x-\frac{qB}{2c}y$. [See (\ref{2.19}) for the definition of $\Pi_x$ in
an arbitrary gauge.] Now identifying the eigenvalue of $\Pi_x$ with $\calp_x$,
those eigenstates are simply
\be        \label{4.5}
\vph_{n,\calp_x,\calp_z}^S(\vr)=e^{i\frac{qB}{2\hbar c}xy}
  \vph_{n,\calp_x,\calp_z}^L(\vr)\,,
\ee
where, for $\vph_{n,\calp_x,\calp_z}^L(\vr)$, one may substitute (\ref{4.3})
(if $\vE=0$) or the expression in (\ref{4.4}) (if $\vE\neq0$). Also, readers
interested in Green's function should consult the Appendix.

We now turn to the two-body system which is described by the equations of
motion (\ref{3.4}). With the harmonic interparticle potential, the appropriate
Hamiltonian is
\be   \label{4.6}
H=\frac1{2m_1}\left[\vp_1-\frac{q_1}c\vA(\vr_1)\right]^2
 +\frac1{2m_2}\left[\vp_2-\frac{q_2}c\vA(\vr_2)\right]^2
 +\frac{k}2|\vr_1-\vr_2|^2-q_1Ey_1-q_2Ey_2.
\ee
The charges $q_1$ and $q_2$ are assumed to satisfy the same conditions as in
Section III. First, consider the case of
$\frac{q_1}{m_1}=\frac{q_2}{m_2}=\frac{Q}{M}$. Then, working in the symmetric
gauge and setting the electric field to zero temporarily, the Hamiltonian
(\ref{4.6}) can be rewritten using the center of mass and relative coordinates
as
\be
H=\frac1{2M}\left[\vP+\frac{Q}{2c}\vR\times\vB\right]^2
+\frac1{2\m}\left[\vp+\frac{m_1m_2Q}{2cM^2}\vr\times\vB\right]^2+\frac12k\vr^2,
\ee
where $\vP$ and $\vp$ are appropriate conjugate momenta:
\be
\vP=\vp_1+\vp_2\,,\quad\qquad
\vp=\frac{m_2}{m_1+m_2}\vp_1-\frac{m_1}{m_1+m_2}\vp_2\,.
\ee
We can construct the energy eigenfunctions in terms of the product states
$\vph(\vR,\vr)=\vph_1(\vR)\vph_2(\vr)$, and that with energy eigenvalue
$\cale=\cale_1+\cale_2$ is obtained if $\vph_1(\vR)$ and $\vph_2(\vr)$
satisfy the eigenvalue equations
\begin{mathletters}
\bea
&&\frac1{2M}\left[\frac\hbar{i}\frac\pa{\pa\vR}
  +\frac{Q}{2c}\vR\times\vB\right]^2\vph_1(\vR)
  =\cale_1\vph_1(\vR)\,,                     \label{4.9a}\\
&&\left\{\frac1{2\m}\left[\frac\hbar{i}\frac\pa{\pa\vr}
  +\frac{m_1m_2Q}{2cM^2}\vr\times\vB\right]^2+\frac12k\vr^2\right\}\vph_2(\vr)
  =\cale_2\vph_2(\vr)\,.                     \label{4.9b}
\eea
\end{mathletters}
Solving (\ref{4.9b}) in the same way as the one-particle Landau-level problem
is solved in the symmetric gauge, one finds the eigenvalue spectrum
\be
\cale_2=\hbar\sqrt{\frac{k}\m}\left(n_z+\frac12\right)
  -\hbar\frac{QB}{2Mc}(n-s)+\hbar\t\o_0(n+s+1)
\ee
(here $\t\o_0=\sqrt{\frac{k}\m+(\frac{QB}{2Mc})^2}$, and $n_z$, $n$ and $s$ can
be arbitrary non-negative integers), with the corresponding eigenfunction
given as ($w=x+iy$)
\be   \label{4.11}
\vph_2(\vr)=\mbox{(const.)}e^{-\frac{\sqrt{\m k}}{2\hbar}z^2}
          H_n\!\left(\left(\frac{\sqrt{\m k}}{\hbar}\right)^\frac12z\right)
          e^{\frac{\m\t\o_0}{2\hbar}\bar ww}\left(\frac\pa{\pa\bar w}\right)^n
          \left(\frac\pa{\pa w}\right)^se^{-\frac{\m\t\o_0}\hbar\bar ww}.
\ee
In the case of identical particles with $q_1=q_2$ and $m_1=m_2$, appropriate
symmetry conditions should be further satisfied by $\vph_2(\vr)$. On the other
hand, the eigenvalue equation (\ref{4.9a}) is precisely the one relevant for
the one-particle Landau-level problem in the symmetric gauge and hence we know
the solutions already, i.e., (\ref{4.5}) but for the reidentification of the
variables involved. Turning on the electric field $\vE=E\hy$ is now
trivial---only the center of mass dynamics, described by the function
$\vph_1(\vR)$ and eigenvalue $\cale_1$, gets affected and the change is
precisely in the same way as in the one-particle problem discussed above.
Analogous analysis can be carried out adopting the Landau gauge also.

We now move on to the case of $q_1=-q_2=q$. For this net neutral two-body
system, the situation becomes more complex. We shall work in the symmetric
gauge, and then the Hamiltonian (\ref{4.6}) (with $\vE=0$) can be expressed
as
\bea  \label{4.12}
H=&&\frac1{2m_1}\left[\vp+\frac{m_1}M\vP
   +\frac{q}{2c}\left(\vR+\frac{m_2}M\vr\right)\times\vB\right]^2\no\\
   &&+\frac1{2m_2}\left[-\vp+\frac{m_2}M\vP
   -\frac{q}{2c}\left(\vR-\frac{m_1}M\vr\right)\times\vB\right]^2
   +\frac12k\vr^2
\eea
Here we further introduce the operators
\be  \label{4.13}
\vPP\eq\vP-\frac{q}{2c}\vr\times\vB\,,\qquad\quad
\vl\eq\vp+\frac{q}{2c}\vR\times\vB\,,
\ee
and then it is easy to show that $(\vPP,\vR)$ and $(\vl,\vr)$ satisfy the
canonical commutation relations:
\bea  \label{4.14}
&&[\Pi_i,X_j]=[l_i,x_j]=-i\hbar\d_{ij}\,,\hspace{7mm}
  [\Pi_i,x_j]=[l_i,X_j]=0\,,\no\\
&&[\Pi_i,l_j]=[\Pi_i,\Pi_j]=[l_i,l_j]=0\,.
\eea
Employing these new variables, the Hamiltonian (\ref{4.12}) reads
\be  \label{4.15}
H=\frac1{2m_1}\left[\vl+\frac{q}{2c}\vr\times\vB+\frac{m_1}M\vPP\right]^2
 +\frac1{2m_2}\left[\vl-\frac{q}{2c}\vr\times\vB-\frac{m_2}M\vPP\right]^2
 +\frac12k\vr^2.
\ee
Evidently, we have $[\Pi_i,H]=0$; $\Pi_i$ are the conserved translation
generators, and all three components are simultaneously diagonalizable (for
the present net neutral system) together with $H$. Adopting the differential
operator realizations $\vPP=\frac\hbar{i}\frac\pa{\pa\vR}$ and
$\vl=\frac\hbar{i}\frac\pa{\pa\vr}$ (which are unitarily equivalent to the
realizations based on $\vP=\frac\hbar{i}\frac\pa{\pa\vR}$ and
$\vp=\frac\hbar{i}\frac\pa{\pa\vr}$), we may thus look for the energy
eigenfunctions having the form
\be
\vph(\vR,\vr)=e^{\frac{i}{\hbar}\vscp\cd\vR}\vph(\vr;\vcp)
\ee
with $\vph(\vr;\vcp)$ satisfying the Schr\"{o}dinger equation appropriate to a
one-body problem:
\bea  \label{4.17}
&&\left\{\frac1{2m_1}\left[\frac\hbar{i}\frac\pa{\pa\vr}
 +\frac{q}{2c}\vr\times\vB+\frac{m_1}M\vcp\right]^2
 +\frac1{2m_2}\left[\frac\hbar{i}\frac\pa{\pa\vr}
 -\frac{q}{2c}\vr\times\vB-\frac{m_2}M\vcp\right]^2\right.\no\\
&&\hspace{20mm}
 +\frac12k\vr^2\Biggr\}\vph(\vr;\vcp)=\cale\vph(\vr;\vcp)\,.
\eea

To solve (\ref{4.17}), we utilize new variables,
$\vr'=\vr-\frac{q}{cM}\frac{\vscp\times\vB}{k+(q^2B^2/c^2M)}$. Note that
the shifted relative position, $\vr'$, entered our classical discussion
also (see (\ref{3.13})). Then, after some straightforward rearrangements,
it is possible to recast (\ref{4.17}) into the form
\bea
&&\left\{\frac1{2\m}\left[\frac\hbar{i}\frac\pa{\pa\vr'}
 +\frac{m_2-m_1}M\frac{q}{2c}\vr'\times\vB
 -\frac{m_2-m_1}{2M}\frac{k}{k+\frac{q^2B^2}{c^2M}}\vcp_\perp\right]^2
 +\frac12\left(k+\frac{q^2B^2}{c^2M}\right){\vr'_\perp}^2\right.\no\\
 &&\hspace{20mm}
 \left.+\frac12k{z'}^2+\frac{k\vcp_\perp^2}{2M(k+\frac{q^2B^2}{c^2M})}
  +\frac{\calp_z^2}{2M}\right\}\vph=\cale\vph\,,
\eea
where $\vr'\eq(\vr'_\perp,z')$ and $\vcp\eq(\vcp_\perp,\calp_z)$. We
further write our wave function as
\be
\vph=e^{\frac{i}\hbar
   \frac{m_2-m_1}{2M}\frac{k}{k+(q^2B^2/c^2M)}\vscp_\perp\cd\vr'_\perp}\t\vph,
\ee
so that the resulting equation for $\t\vph$ may assume the (almost) same form
as (\ref{4.9b}):
\bea  \label{4.20}
\left\{\frac1{2\m}\left[\frac\hbar{i}\frac\pa{\pa\vr'}
 +\frac{m_2-m_1}M\frac{q}{2c}\vr'\times\vB\right]^2
 +\frac12\left(k+\frac{q^2B^2}{c^2M}\right){\vr'_\perp}^2+\frac12k{z'}^2
 \right\}\t\vph=\t\cale\t\vph,\no\\
 \left(\t\cale=\cale-\frac{k}{k+\frac{q^2B^2}{c^2M}}\frac{\vcp_\perp^2}{2M}
 -\frac{\calp_z^2}{2M}\right).
\eea
The eigenfunctions of (\ref{4.20}) are thus given by the expression
(\ref{4.11}) (with $(x,y,z)$ taken by $(x',y',z')$) if the parameter $\t\o_0$
there is suitably adjusted; in the present case, $\t\o_0$ should be identified
with $\sqrt{\frac1\m(k+\frac{q^2B^2}{c^2M})
+\frac{q^2B^2(m_1-m_2)^2}{4c^2m_1^2m_2^2}}$. [Recall that this frequency also
figured in in our classical discussion.] This eigenfunction, which is
denoted as $\t\vph_{n_z,n,s}(\vr')$, corresponds to the eigenvalue
\be
\t\cale=\hbar\sqrt{\frac{k}\m}\left(n_z+\frac12\right)
       -\hbar\frac{(m_2-m_1)qB}{2m_1m_2c}(n-s)+\hbar\t\o_0(n+s+1).
\ee
Note that the functions $\t\vph_{n_z,n,s}(\vr')$ have no dependence on $\vcp$
except through the definition of $\vr'$.

The eigenfunctions found above are the ones appropriate when $\vPP$ and $\vl$
are realized as $\frac\hbar i\frac\pa{\pa\vR}$ and $\frac\hbar
i\frac\pa{\pa\vr}$. To make direct connection with our discussion in Section
II, it should be desirable to have them brought to the expressions appropriate
to more conventional realizations $\vP=\frac\hbar{i}\frac\pa{\pa\vR}$ and
$\vp=\frac\hbar{i}\frac\pa{\pa\vr}$. As one can easily verify, this job is
effected by a simple multiplicative phase,
$e^{\frac{i}\hbar\frac{q}{2c}\vR\cd\vr\times\vB}$. Including this phase factor,
we may now express the complete eigenfunctions of the Hamiltonian (\ref{4.15})
by
\be   \label{4.22}
\vph_{\vscp,n_z,n,s}(\vR,\vr)
=e^{\frac{i}\hbar\frac{q}{2c}\vR\cd\vr\times\vB}e^{\frac{i}\hbar\vscp\cd\vR}
 e^{\frac{i}\hbar\frac{m_2-m_1}{2M}\frac{k}{k+(q^2B^2/c^2M)}\vscp_\perp\cd
    \vr_\perp}
 \left.\left[\t\vph_{n_z,n,s}(\vr')\right]
 \right|_{\vr'=\vr-\frac{q}{cM}
  \frac{\vsscp_\perp\times\vB}{k+(q^2B^2/c^2M)}}
\ee
with the energy eigenvalue given as
\bea   \label{4.23}
\cale_{\vscp,n_z,n,s}&=&\frac{k}{k+\frac{q^2B^2}{c^2M}}\frac{\vcp_\perp^2}{2M}
 +\frac{\calp_z^2}{2M}\no\\
 &&+\hbar\sqrt{\frac{k}{\m}}\left(n_z+\frac12\right)
 +\hbar\t\o_0(n+s+1)-\hbar\frac{(m_2-m_1)qB}{2m_1m_2c}(n-s)\\
 &&\hspace{47mm}(\mbox{$n_z$, $n$, $s$: non-negative integers}).\no
\eea
For these energy eigenstates, the expectation value of the relative position
operator $\vr=\vr_1-\vr_2$ is consistent with the classical result in
(\ref{3.16}). Are the expressions (\ref{4.22}) also consistent with the
general form of the eigenfunctions we have in (\ref{2.18})? Yes, indeed. For
that, it suffices to notice that the first phase factor in the right hand side
of (\ref{4.22}) can be rewritten as
\be  \label{4.23a}
e^{\frac{i}\hbar\frac{q}{2c}\vR\cd\vr\times\vB}
 =e^{i\frac{qB}{2\hbar c}(x_1y_1-x_2y_2)}
  e^{-i\frac{qB}{\hbar c}xY-i\frac{qB}{2\hbar c}\frac{m_2-m_1}Mxy}.
\ee
We use this in (\ref{4.22}) and then, except for the appropriate gauge
transformation factor $e^{i\frac{qB}{2\hbar c}(x_1y_1-x_2y_2)}$ (needed to
convert the Landau-gauge results into the symmetric-gauge results), the
dependence on the variable $X$ in the resulting expression is entirely in
$e^{\frac{i}\hbar\calp_xX}$. Hence ours are fully consistent with
(\ref{2.18}).

Energy eigenstates in the presence of the additional electric field $\vE=E\hy$
can readily be identified also. In the symmetric gauge, our recipe tells us
that the energy eigenfunctions for this case are
\bea   \label{4.24}
\vph'_{\vscp,n_z,n,s}(\vR,\vr)
=e^{\frac{i}\hbar\frac{q}{2c}\vR\cd\vr\times\vB}e^{\frac{i}\hbar\vscp\cd\vR}
 e^{\frac{i}\hbar\frac{m_2-m_1}{M}\frac{k}{k+(q^2B^2/c^2M)}\vscp^*_\perp\cd
    \vr_\perp}
 &&\left.\left[\t\vph_{n_z,n,s}(\vr')\right]
 \right|_{\vr'=\vr-\frac{q}{cM}
  \frac{\vsscp^*_\perp\times\vB}{k+(q^2B^2/c^2M)}},\no\\
&&\left(\vcp^*_\perp\eq\vcp_\perp-\frac{cE}{B}M\hx\right)
\eea
with the energy eigenvalue given by the formula
$\cale'_{\vscp,n_z,n,s}
 =[\cale_{\vscp,n_z,n,s}]_{\vscp\ra(\vscp^*_\perp,\calp_z)}
  +\frac{cE}B\calp_x-\frac12M\frac{c^2E^2}{B^2}$
($\cale_{\vscp,n_z,n,s}$ is in (\ref{4.23})). Note that $\vcp$ in
(\ref{4.24}) corresponds to the eigenvalue of the translation generator $\vPP$
(see (\ref{4.13})). In this gauge the center of mass velocity operator is
$\dot\vR=\frac1M[\vPP+\frac{q}c\vr\times\vB]$, and hence for the eigenstate
(\ref{4.24}) its expectation value is found to be
\bea  \label{4.25}
\bra\dot\vR\ket&=&\frac1M\left[\vcp+\frac{q}c\bra\vr\ket\times\vB\right]\no\\
  &=&-\frac{kc}{qB^2}\bra\vr\ket\times\vB+\frac{cE}B\hx+\frac{\calp_z}M\hz\,,
\eea
where, on the second line, we have dispensed with $\vcp_\perp$ using the
formula
$\bra\vr\ket=\frac{q}{cM}\frac{\vscp^*_\perp\times\vB}{k+(q^2B^2/c^2M)}$. The
relation (\ref{4.25}) has the classical counterpart in (\ref{3.17}).

In principle, it should be possible to perform an analogous analysis for a
three-body system also, say, assuming charge values $(2q, -q, -q)$ and mutual
harmonic interactions between the particles. This could give some
rough information on the behavior of a neutron in the presence of constant
electromagnetic fields.

\section{DISCUSSION AND SUMMARY}
In this paper we discussed the physical role of Galilean transformation in
general non-relativistic charged-particle systems in the presence of a uniform
magnetic field. By a Galilean transform with a judicious boost velocity, one
obtains the same system but now under mutually perpendicular magnetic and
electric fields. Thus any problem under the latter circumstance is completely
solved in terns of a judicious Galilean transform of the same problem under a
uniform magnetic field only. Applications of this observation have been made
for relatively simple, classical and quantum mechanical, systems. This
observation is clearly at the heart of the Hall effect, and also explains the
Stark-like effect exhibited by a neutral composite system in motion under a
uniform magnetic field.

Galilean transformations can be applied to the anyon system in a uniform
magnetic field. [See Ref. 9 for related literatures.] A particularly convenient
way to deal with the unconventional statistical property of anyons is by
introducing the Aharonov-Bohm-type interactions for every pair of particles.
These Aharonov-Bohm interactions are still Galilean invariant, and hence our
discussion given in Section II requires no essential change. Consequently, once
one has solutions to the problem of anyons in the presence of a
uniform magnetic field, those results may simply be Galilean-transformed
(according to the procedures detailed by us in Section II) to obtain
corresponding solutions in the presence of mutually perpendicular magnetic and
electric fields. In this aspect, anyons are not different from ordinary
particles.

\acknowledgments
This work was supported in part by the Korea
Science and Engineering Foundation (through the Center for Theoretical
Physics, Seoul National University) and also by the Ministry of Education,
Korea. One of us (C.K.) wishes to thank the Center for Theoretical Physics at
MIT, where this work was partially undertaken, and the NSF-KOSEF exchange
program which made this visit possible. We acknowledge some useful comments
from R. Jackiw and S. K. Kim.

\appendix
\def\rfi{(\vr_f,t_f;\vr_i,t_i)}
\renewcommand{\theequation}{A.\arabic{equation}}
\section*{GALILEAN TRANSFORMATION OF GREEN'S FUNCTION}
Let $G'\rfi$ denote the quantum mechanical Green's function in the one-particle
system defined by the Hamiltonian (\ref{4.1}), and $G\rfi$ Green's function in
the presence of the magnetic field $\vB$ only, i.e., $G\rfi=[G'\rfi]|_{E=0}$.
Then, looking at (\ref{2.8}) and (\ref{2.11}), astute readers will immediately
make the identification
\bea
&&G'\rfi=e^{\frac{i}{\hbar}[\a(\vr_f,t_f)-\a(\vr_i,t_i)]}
  G(\vr_f+\vu t_f,t_f;\vr_i+\vu t_i,t_i),            \label{A.1}\\
&&\a\rt=-m\vu\cd\vr-\frac12m\vu^2t-\frac{q}c\{\t\LLL(\vr)-\t\LLL(\vr+\vu t)\},
\eea
where $\vu\eq-c\frac{\vE\times\vB}{B^2}$, and $\t\LLL(\vr)$ is defined through
(\ref{2.9}). Hence, given the explicit form for $G\rfi$, the expression for
$G'\rfi$ may also be written down using this relation. Moreover, if
$L'(\vr,\dot\vr)$ denotes the Lagrangian corresponding to the Hamiltonian
(\ref{4.1}) and $L(\vr,\dot\vr)$ that in the presence of the $\vB$-field only,
it is possible to demonstrate by direct calculation that
\be
L'(\vr,\dot\vr)=L(\vr+\vu t,\dot\vr+\vu)+\frac{d}{dt}\a\rt\,,
\hspace{10mm}\left(\vu=-c\frac{\vE\times\vB}{B^2}\right).
\ee
By considering this relation together with the path integral representation of
Green's function, one has an alternative understanding of the above
relationship between the Green's functions.

Explicitly, in the Landau gauge, we have
\bea
G\rfi=&&\th(t_f-t_i)\left(\frac{m}{2\pi i\hbar(t_f-t_i)}\right)^{\frac32}
  \left(\frac{\frac{\o_c}2(t_f-t_i)}{\sin[\frac{\o_c}2(t_f-t_i)]}\right)\no\\
  &&\times e^{\frac{im}{2\hbar}\{\frac{(z_f-z_i)^2}{t_f-t_i}
    +\frac{\o_c}2\cot[\frac{\o_c}2(t_f-t_i)][(x_f-x_i)^2+(y_f-y_i)^2]
    +\o_c(x_f-x_i)(y_i+y_f)\}}
\eea
Then, using (\ref{A.1}), we can immediately write corresponding Green's
function in the presence of both $\vB=B\hz$ and $\vE=E\hy$ as (here,
$\vu\eq-c\frac{E}B\hx$)
\bea
G'\rfi=&&\th(t_f-t_i)\left(\frac{m}{2\pi i\hbar(t_f-t_i)}\right)^{\frac32}
      \left(\frac{\frac{\o_c}2(t_f-t_i)}{\sin[\frac{\o_c}2(t_f-t_i)]}\right)
        e^{-imu(x_f-x_i)-\frac{i}2mu^2(t_f-t_i)}\no\\
 &&\hspace{-12mm}\times e^{\frac{im}{2\hbar}\{\frac{(z_f-z_i)^2}{t_f-t_i}
  +\frac{\o_c}2\cot[\frac{\o_c}2(t_f-t_i)][(x_f-x_i+u(t_f-t_i))^2+(y_f-y_i)^2]
  +\o_c[x_f-x_i+u(t_f-t_i)](y_i+y_f)\}}.\no\\
 &&
\eea

\end{document}